\documentclass[%
preprint,
superscriptaddress,
amsmath,amssymb,
aps,prl,floatfix
]{revtex4-1}

\usepackage{booktabs}
\usepackage{tabularx}
\usepackage{braket}
\usepackage{float}
\usepackage{dcolumn}
\usepackage{bm}
\usepackage{subfigure}
\usepackage{color}

\usepackage{graphicx,amsmath,amssymb,bm,epstopdf,epsfig,psfrag,rotate}
\graphicspath{{pic/}}

\newcommand\uu{{\bf u}}
\newcommand\dd{{\partial }}

\newcommand\Nu{\nobreak\mbox{$\mathcal N$\hskip-0.95mm$u$}}

\newcommand\Ra{\nobreak\mbox{$\mathcal R$\hskip-0.3mm$a$}}

\newcommand\Rey{\nobreak\mbox{$\mathcal R$\hskip-0.3mm$e$}}
\newcommand\re{\nobreak\mbox{$\mathcal R$\hskip-0.3mm$e$}}

\newcommand\Pran{\nobreak\mbox{$\mathcal P$\hskip-0.3mm$r$}}
\newcommand\pr{\nobreak\mbox{$\mathcal P$\hskip-0.3mm$r$}}

\newcommand{\oo}{\color{black} \normalfont}
\newcommand{\bb}{\color{black} \normalfont}

\def\gtwid{\mathrel{\raise.3ex\hbox{$>$\kern-.75em\lower1ex\hbox{$\sim$}}}}
\def\ltwid{\mathrel{\raise.3ex\hbox{$<$\kern-.75em\lower1ex\hbox{$\sim$}}}}

\def\agt{\mathrel{\raise.3ex\hbox{$>$\kern-.75em\lower1ex\hbox{$\sim$}}}}
\def\alt{\mathrel{\raise.3ex\hbox{$<$\kern-.75em\lower1ex\hbox{$\sim$}}}}

\def\nn{{\mbox{\boldmath$\nabla$}}}

\def\be{\begin{equation}}
\def\ee{\end{equation}}
\newcommand{\ba}{\begin{eqnarray}}
\newcommand{\ea}{\end{eqnarray}}

\newcommand{\bea}{\begin{eqnarray}}
\newcommand{\eea}{\end{eqnarray}}
\newcommand{\bean}{\begin{eqnarray}}
\newcommand{\eean}{\end{eqnarray}}

\definecolor{blue5}{RGB}{0, 13, 52}
\definecolor{blue4}{RGB}{0, 102, 155}
\definecolor{blue3}{RGB}{0, 137, 204} 
\definecolor{blue2}{RGB}{7, 170, 255}
\definecolor{blue1}{RGB}{130, 211, 255}
\definecolor{pink1}{RGB}{255, 117, 186}
\definecolor{pink2}{RGB}{255, 0, 124}
\definecolor{pink3}{RGB}{196, 0, 96} 
\definecolor{pink4}{RGB}{153, 0, 76}
\definecolor{pink5}{RGB}{52, 0, 13}
\definecolor{pink5}{RGB}{52, 0, 13}

\definecolor{gfblue4}{RGB}{0, 102, 155}
\definecolor{gfred4}{RGB}{153, 0, 76}

\begin{document}

\title{Ultimate regime of 
Rayleigh--B\'enard turbulence:\\   Sub-regimes and their scaling relations for $\Nu$ vs. $\Ra$ and $\pr$}


\author{Olga~Shishkina}
\email[]{Olga.Shishkina@ds.mpg.de}
\affiliation{Max Planck Institute for Dynamics and Self-Organization, 37077 G\"ottingen, Germany}
\author{Detlef~Lohse}
\email[]{d.lohse@utwente.nl}
\affiliation{Physics of Fluids Department, J.M. Burgers Center for Fluid Dynamics, 
and Max Planck -- University of Twente Center for Complex Fluid Dynamics;
Faculty of Science and Technology, 
University of Twente, Enschede, The Netherlands}
\affiliation{Max Planck Institute for Dynamics and Self-Organization, 37077 G\"ottingen, Germany}

\date{\today}


\begin{abstract}
We offer a new model for the heat transfer and the turbulence intensity in strongly driven Rayleigh--B\'enard turbulence (the so-called ultimate regime), which in contrast to hitherto models is consistent with the new mathematically exact 
heat transfer upper bound of 
Choffrut {\it et al.} [J. Differential Equations {\bf 260}, 3860 (2016)]
 and thus enables extrapolations of the heat transfer to geo- and astrophysical flows. 
The model  distinguishes between four subregimes of the ultimate regime  and well describes the measured heat transfer in various large-$\Ra$ experiments. In this new representation, which properly accounts for the Prandtl number dependence, the onset to the ultimate regime is seen in all available large-$\Ra$ data sets, though at different Rayleigh numbers, as to be expected for a non-normal--nonlinear instability. 
\end{abstract}

 \maketitle
 
Knowing the heat and/or mass transfer in large-scale turbulent flows is of utmost importance for many questions in climate research, in geophysical or astrophysical systems, or in industrial flow systems. 
Examples are  thermally driven flows in the ocean, in  the atmosphere \cite{Vallis2017}, or in the outer core of Earth, other planets, or stars \cite{Clarke2007}. In all these cases, very strong turbulence is  achieved, due to the strong thermal driving. 
For such systems, however, direct measurements under controlled conditions are not feasible, and neither are direct numerical simulations,  given the many degrees of freedom of such systems, though the underlying dynamical equations (the  advection-diffusion equations for the temperature and/or the mass transport, coupled to the Navier--Stokes equations) are known. Given this, in order to get an estimate for the heat and/or mass transfer in such systems, one has to rely on more controlled model systems on much smaller scale and then extrapolate towards larger systems with stronger thermal driving.
 
The most popular controlled model system for heat transfer is the Rayleigh--B\'enard (RB) system, consisting of a container of height $L$ filled with fluid, heated from below and cooled from above 
 \cite{Chandrasekhar1961,Bodenschatz2000,Kadanoff2001,Ahlers2009,Lohse2010,Chilla2012,Xia2013,Shishkina2021,Lohse2023}. 
The control parameters of this thermally driven convective flow are the Rayleigh number $\Ra$ (the dimensionless temperature difference $\Delta$ between top and bottom plates, as measure of the driving strength), the Prandtl number $\pr$ (the ratio between kinematic viscosity $\nu$ and thermal diffusivity $\kappa$), and the aspect ratio $\Gamma$
 (the width 
of the system divided by its height). 
The main global response parameters  are the Nusselt number $\Nu$ (the dimensionless heat transfer from bottom to top) and the so-called wind Reynolds number $\re$, which quantifies the velocity of the large scale convective flow. 
The key question is: How do the Nusselt and the Reynolds number depend on the control parameters, $\Nu (\Ra, \pr, \Gamma )$ and $\re (\Ra, \pr, \Gamma )$? 
For not too strong thermal driving (the so-called classical regime) this question can meanwhile be answered and there is 
good agreement between various experiments and numerical simulations and a good understanding of the flow physics, namely in terms of the ``Grossmann--Lohse-theory'' or in short ``GL-theory'', cf.\  \cite{Grossmann2000,Grossmann2001,Grossmann2002,Stevens2013}, see also the reviews \cite{Ahlers2009,Lohse2023}.  
 
This is not so for very strong thermal driving, i.e., for 
the regime of
very large Rayleigh numbers, which is called the ``ultimate regime'' and to which, for large enough $\Ra$, the RB system is believed to undergo a transition of non-normal--nonlinear type \cite{Roche2020,Lohse2023}, as typical in strongly driven  sheared wall-bounded flows \cite{Avila2023} and as here is the case in the boundary layers.  
It 
is this very ultimate regime which is relevant for climate research and the above mentioned geophysical and astrophysical convective flows, due to the very strong thermal driving in these systems. 
Therefore, extrapolations from the classical regime to the ultimate regime are required.
Typically, these extrapolations are
sought for as scaling laws, but 
this only makes sense once there is no transition towards a different state of turbulence. 
If there is such a transition, the  extrapolation with a simple scaling law becomes meaningless. 
But then, how to upscale the RB system and how to understand and predict the heat (and mass) flux for very large $\Ra$, as it occurs in geo- and astrophysical applications? 

To answer these questions, various theoretical heuristic models of different degrees  of rigor have been developed, based on some assumptions and speculations on the flow physics in this ultimate regime \cite{Priestley1954,Malkus1954,Kraichnan1962,Herring1963,Stewartson1966,Roberts1966,Spiegel1971,Castaing1989,Shraiman1990,Chavanne1997,Grossmann2011,Grossmann2012}.  
The most famous and influential one of all these models may be  the one by Kraichnan \cite{Kraichnan1962}, who for
very large $\Ra$ and small $\pr \le 0.15$ suggested 
$\Nu \sim \Ra^{1/2} \pr^{1/2} / (\log \Ra )^{3/2}$. For very large $\Ra$ and moderate $0.15 < \pr \le 1$,  he suggested a slightly different 
$\pr$-dependence, namely 
$\Nu \sim \Ra^{1/2} \pr^{-1/4} / (\log \Ra )^{3/2}$. 

All these models obviously should obey the mathematically strict upper bounds for $\Nu (\Ra, \pr )$, which can be derived from the
underlying dynamical equations (heat transfer equation and Navier--Stokes equations in their Boussinesq approximation).
The well-known upper bound  $\Nu<A\,\Ra^{1/2}$ was found already in the second half of the last century 
\cite{Howard1963,Busse1969,Howard1972,Doering1996}
and the best-known (smallest) prefactor $A\approx0.026$ was calculated in \cite{Plasting2003}.
Although this upper bound leads to values much higher than the experimentally measured or numerically calculated $\Nu$,
it excludes the universality  of the scaling relation $\Nu\sim\Ra^{1/2}\Pran^{1/2}$ 
 (with any logarithmic corrections).  This scaling relation  was
 proposed in several models for the ultimate regime \cite{Grossmann2011,Chavanne1997,Spiegel1971}, but 
 due to the upper bound 
 it   can only hold for $\Pran^{1/2}\lesssim A$.
Moreover, in 2016  Choffrut {\it et al.} \cite{Choffrut2016} succeeded to sharpen the upper bound in a large-$\Pran$ subrange of the ultimate regime, namely to $\Nu \lesssim\Ra^{1/3}$ for $\Pran\gtrsim\Ra^{1/3}$ (all subject to logarithmic corrections). 
Thus in this subregime of the ultimate regime, in which  $\Pran$ grows faster than $\Ra^{1/3}$ but slower than $\Ra^{2/3}$ (that is, as $\Pran\sim\Ra^{a}$, $1/3<a<2/3$), Kraichnan's model predicts $\Nu\sim\Ra^{\gamma}$ with $\gamma=-a/4+1/2>1/3$, in direct contradiction to Choffrut {\it et al.}'s strict upper bound.
Similarly, also other models \cite{Grossmann2011,Chavanne1997,Spiegel1971}, which propose the growth of $\Nu$ faster than $\Ra^{1/3}$ for the ultimate regime, cannot hold in this subrange of the $\Ra-\Pran$ parameter space. 

This discrepancy calls for revisiting the suggested scaling laws in the ultimate regime, in view of the improved and sharpened strict upper bound \cite{Choffrut2016}. 
To do so is the objective of this paper. 
We will first suggest a new model for the heat transfer in the ultimate regime, which is based on the flow physics in the turbulent boundary layers and which respects the new mathematically strict upper bounds in the various subregimes of the ultimate regime  in the $\Ra-\Pran$ parameter space. 
We will then show that the available experimental and numerical data for strongly driven RB convection can be well described with our new model. 

\oo
We start with the boundary layer equations for the horizontal velocity $u_x$ and temperature $\theta$ in a turbulent flow next to a rigid horizontal wall:
\ba
\label{m1}
\partial_t u_x + {\uu}\cdot\nn u_x +\dd_x p/\rho&=& \nu \nn^2 u_x,\\
\label{m2}
\partial_t \theta + {\uu}\cdot\nn \theta &=& \kappa \nn^2\theta
\ea
(where $t$ denotes time, $x$ and $z$, respectively, the horizontal and vertical coordinates, $p$ the hydrodynamic pressure, and $\rho$ the density)
and conduct the Reynolds decomposition of the flow components into their time-averages and fluctuations:
$\uu = \langle\uu\rangle_t + \uu'$, $\theta = \langle\theta\rangle_t + \theta'$.
We assume that the flow is highly turbulent, so that the convective contributions from the mean, time-averaged $\langle*\rangle_t$ flow are negligible compared to the contributions from the Reynolds stresses, i.e.   
$\left|\langle\uu\rangle_t \cdot \langle \nn u_x\rangle_t\right|\ll \left|\langle \uu'\cdot\nn u_x'\rangle_t\right|$
and
$\left|\langle\uu\rangle_t \cdot \langle \nn\theta\rangle_t\right|\ll \left| \langle \uu'\cdot\nn \theta'\rangle_t\right|$, so that the following relations hold:
$\langle \uu\cdot\nn u_x\rangle_t\approx\langle \uu'\cdot\nn u_x'\rangle_t=\nn\cdot\langle u_x'\uu'\rangle_t$,
$\langle \uu\cdot\nn \theta\rangle_t\approx\langle \uu'\cdot\nn \theta'\rangle_t=\nn\cdot\langle \theta'\uu'\rangle_t$.
Using these relations, we average Eqs.~(\ref{m1}), (\ref{m2}) in time and over a horizontal cross-section $S$ (i.e., apply $\langle*\rangle_{t, S}$)
under further natural assumptions that the long averages in time of the temporal derivatives vanish,
$\langle\dd_t u_x\rangle_{t, S}=0$,
$\langle\dd_t \theta\rangle_{t, S}=0$,
as well as the averages in the horizontal direction of the horizontal derivatives,
$\langle\dd^2_x u_x\rangle_{t, S}=0$,
$\langle\dd^2_x \theta\rangle_{t, S}=0$,
$\langle\dd_x p\rangle_{t, S}=0$,
$\nn\cdot\langle u_x'\uu'\rangle_{t, S}=\dd_z\langle u_z'u_x'\rangle_{t, S}$,
$\nn\cdot\langle \theta'\uu'\rangle_{t, S}=\dd_z\langle u_z'\theta'\rangle_{t, S}$.
With this we obtain the following reduced equations:
\ba
\label{m3}
\partial_z \langle u_z'  u_x'\rangle_{t, S} &=& \nu  \partial^2_z \langle u_x\rangle_{t, S},\\
\label{m4}
\partial_z \langle u_z'  \theta'\rangle_{t, S} &=& \kappa  \partial^2_z \langle\theta\rangle_{t, S}.
\ea
Integrating equations (\ref{m3})--(\ref{m4}) from 0 to $z$, introducing the eddy viscosity $\nu_{\tau}$ and 
the eddy thermal diffusivity $\kappa_{\tau}$,
\ba
\label{m5}
\langle u_z' u_x' \rangle_{t, S} &\equiv& -\nu_{\tau} \partial_z \langle u_x\rangle_{t, S},\\
\label{m6}
\langle u_z' \theta' \rangle_{t, S} &\equiv& -\kappa_{\tau} \partial_z \langle\theta\rangle_{t, S},
\ea
and taking into account the vanishing fluctuations at the plate and that the Nusselt number
is defined  by 
$\Nu\equiv-\left.\dd_z \langle\theta\rangle_{t, S}\right|_{z=0} L/\Delta$, 
we obtain 
\be
\label{m7}
u^2_{\tau} \equiv\left.\nu\dd_z \langle u_x\rangle_{t, S}\right|_{z=0}
= (\nu+\nu_{\tau})\dd_z \langle u_x\rangle_{t, S} 
\ee
for (the square of) the friction velocity and 
\oo
\be
\label{m8}
({\kappa\Delta}/{L})\Nu = -(\kappa+\kappa_{\tau})\dd_z \langle\theta\rangle_{t, S}.
\ee
\bb

To close the system (\ref{m7}) and (\ref{m8}), we need to know the functional dependences of the eddy viscosity $\nu_{\tau}(z)$ and the eddy thermal diffusivity $\kappa_{\tau}(z)$.
Near the plate, within the viscous sublayer of the thickness $z_{\tau}\equiv \nu/u_\tau$, both the eddy viscosity $\nu_{\tau}(z)$ and the eddy thermal diffusivity $\kappa_{\tau}(z)$, behave as cubic functions of the distance from the plate, $\propto z^3$ \cite{Antonia1980,Antonia1991,Shishkina2015,Shishkina2017a}, and therefore the contribution of the eddy viscosity and eddy thermal diffusivity within the viscous sublayer is negligible.

To estimate the mean vertical profiles of $\nu_{\tau}(z)$, $\kappa_{\tau}(z)$ and $\epsilon_u(z)$ outside the viscous sublayer, we follow Landau \citep{Landau1987} and assume that (i) the turbulent Prandtl number $\Pran_{\tau}\equiv \nu_{\tau}/\kappa_{\tau}$ is independent of (or only weakly dependent on) the molecular Prandtl number $\Pran$ and (ii) that the mean vertical profiles of $\nu_{\tau}(z)$, $\kappa_{\tau}(z)$,  and $\epsilon_u(z)$, are determined exclusively by the momentum transferred by the  fluid to a solid wall, i.e. the friction velocity $u_{\tau}$, and by the distance to the plate, subject to a certain Prandtl-number dependence, i.e. $z\Pran^\zeta$.

These assumptions, by dimensional analysis, imply that, outside the viscous sublayer,  $\nu_{\tau}(z)$, $\kappa_{\tau}(z)$, and $\epsilon_u(z)$ should scale as
\oo
\ba
\label{m9}
\nu_{\tau}(z)&=&\varkappa \,u_{\tau}z \,\Pran^\zeta,\\ 
\label{m10}
\kappa_{\tau}(z)&=&\varkappa_{\theta} \,u_{\tau}z \,\Pran^\zeta,\\
\label{m11}
\epsilon_u(z)&=&\dfrac{\varkappa_\epsilon \,u_{\tau}^3}{z \,\Pran^\zeta},
\ea
with some positive constants $\varkappa$, $\varkappa_{\theta}$, and $\varkappa_\epsilon$.
\bb

We further propose that the turbulent diffusivities $\nu_{\tau}$ and $\kappa_{\tau}$ are controlled by the smallest of the two fluid characteristics of diffusion, i.e., either by $\nu$ or by $\kappa$. 
In other words, both, $\nu_{\tau}$ and $\kappa_\tau$, should be proportional to $\sqrt{\left.\nu\dd_z \langle u_x\rangle_t\right|_{z=0}}$ for small $\Pran \lesssim1$, and to $\sqrt{\left.\kappa\dd_z \langle u_x\rangle_t\right|_{z=0}}$ for large $\Pran\gtrsim1$, which implies $\zeta=0$ for $\Pran\lesssim1$ and $\zeta=-1/2$ for $\Pran\gtrsim1$.
\oo
This proposition is related to the fact that if any of the two quantities, $\nu$ or $\kappa$, equals zero, convection is fully suppressed, $\Nu=1$, independently from which value takes the other characteristics of diffusion ($\kappa$ or  $\nu$, respectively).
\bb

\begin{figure}[htb]
\includegraphics[width=0.45\columnwidth]{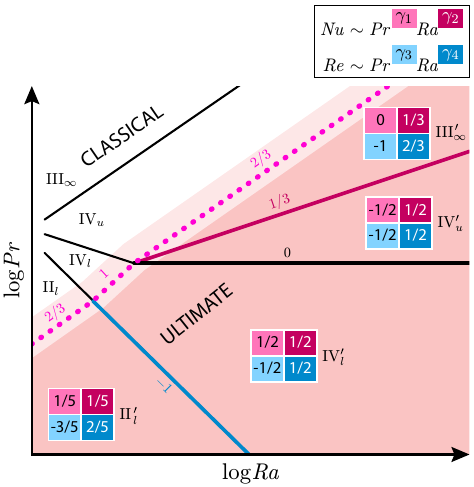}
\caption{A sketch of the proposed scaling relations in the ultimate regime of Rayleigh--B\'enard convection 
in the $\Pran-\Ra$ parameter space, where the ultimate regime is split into the subregimes IV$_u'$, IV$_\ell'$, III$'_\infty$ and II$'_\ell$. 
The numbers in color boxes show the scaling exponents in the relations 
$\Nu\sim\Pran^{^{\colorbox{pink1}{\scriptsize$\gamma_1$}}}\Ra^{^{\colorbox{pink3}{\color{white}\scriptsize$\gamma_2$}}}$,
$\Rey\sim\Pran^{^{\colorbox{blue1}{\scriptsize$\gamma_3$}}}\Ra^{^{\colorbox{blue3}{\color{white}\scriptsize$\gamma_4$}}}$ (subject to logarithmic corrections).
The straight lines indicate the slopes of the transitions between the neighbouring regimes, 
$\Pran\sim\Ra^\eta$, where the values of $\eta$
are written next to the lines.
The dotted line indicates where the laminar kinetic boundary layer is expected to become turbulent (i.e., where
 the shear Reynolds number achieves a critical value, $\Rey_s=const.$).
}
\label{ult} 
\end{figure}

Now, from the exact relation $ \epsilon_u \equiv \nu \langle(\nn\uu)^2\rangle=\nu^3 L^{-4}\Ra\,\Pran^{-2}(\Nu-1)$ for the 
time- and volume-averaged kinetic energy dissipation rate \citep[see, e.g.,][]{Shraiman1990} and Eqs.~(\ref{m7})--(\ref{m11})
we can derive the scaling relations, which we propose for the ultimate regime, 
for both small and large $\Pran$: 
Dividing both sides of Eq.~(\ref{m7}) by $(\nu+\nu_{\tau})$, substituting (\ref{m9}), and integrating the resulting equation in $z$ from the edge of the viscous sublayer, $z_{\tau}\equiv \nu/u_\tau$, to the location $L'\sim L/2$ of the maximal wind velocity  
\oo
$\left.\langle u_x\rangle_{t, S}\right|_{z=L'}  = \nu \re/L$
\bb
we obtain
\ba
{\Rey}&\sim&\frac{\Rey_{\tau}}{\varkappa\,\Pran^{\zeta}}\log\left(\frac{\varkappa}{2}\Rey_{\tau}\Pran^{\zeta}+1\right) \nonumber\\
&\sim&{\Rey_{\tau}\,\Pran^{-\zeta}\log\Rey_{\tau}}. 
\label{m12}
\ea
Analogously, dividing both sides of Eq.~(\ref{m8}) by $(\kappa+\kappa_{\tau})$, substituting (\ref{m10}), and integrating the resulting equation in $z$ from $z_{\tau}$ to $L/2$, we obtain
\be
\label{m13}
{\Nu}\sim\dfrac{\frac{\varkappa_{\theta}}{2}\Rey_{\tau}\Pran^{\zeta+1}}{\log\left(\frac{\varkappa_{\theta}}{2}\Rey_{\tau}\Pran^{\zeta+1}+1\right)}
\sim\;{\dfrac{\Rey_{\tau}\Pran^{\zeta+1}}{\log\Rey_{\tau}}}.
\ee
In Eqs.~(\ref{m12}) and (\ref{m13}) behind the second tilde-sign we have neglected the $\pr$-dependences 
\oo
and keep only the leading terms in  the log-corrections.
Next we consider the profile $\epsilon_u(z)$ of the kinetic energy dissipation rate.  
As $\epsilon_u(0)\sim\nu(\left.\dd_z \langle u_x\rangle_{t, S}\right|_{z=0})^2\sim u_\tau^4/\nu$, 
\bb
the contribution to the mean kinetic energy dissipation rate from the viscous sublayer is smaller than $\epsilon_u(0)z_\tau\sim(u_\tau^4/\nu)(\nu/u_\tau)=u_\tau^3$.
In contrast, the contribution from the core part of the domain is scaling-wise larger, as one can see from integrating (\ref{m11}), 
\oo
\be
\label{m144}
\int_{z_\tau}^{L/2}\epsilon_u(z)dz \sim
{\varkappa_\epsilon u_{\tau}^3\over \Pran^\zeta}\log\left({\Rey_{\tau}}/{2}\right) 
\gtrsim u_{\tau}^3\log \Rey_{\tau}. 
\ee
\bb
Here we used the fact that $\Pran^{-\zeta}\geq1$ for all Prandtl numbers.
Since the main contribution to the total kinetic energy dissipation rate $\epsilon_u$ comes from the bulk, with the exact relation for $\epsilon_u$  we obtain
\be
\label{m14}
{2\over L}\int_{z_\tau}^{L/2}\epsilon_u(z)dz \approx \epsilon_u = \dfrac{\nu^3}{L^{4}}\Ra\,\Pran^{-2}(\Nu-1).
\ee
From relations (\ref{m144}) and (\ref{m14})
it follows
\ba
{\Ra\,\Nu\,\Pran^{-2}}
&\sim&{\Rey_{\tau}^3\, \Pran^{-\zeta}}\log(\Rey_{\tau}).
\label{m15}
\ea
Combining (\ref{m12}), (\ref{m13}) and (\ref{m15}) we obtain
\be
\label{m16}
\Rey\sim\Pran^{-1/2}\Ra^{1/2}  \quad \text{ for all } \;\Pran,
\ee
and 
\be
\label{m166}
\Nu\sim{\Pran^{2\zeta+1/2}\Ra^{1/2}\over(\log\Ra)^2}
\ee
with $\zeta=0$ for $\Pran\lesssim1$ and $\zeta=-1/2$ for $\Pran\gtrsim1$.
Note that in relations (\ref{m16}) and (\ref{m166}) we again neglected the $\pr$-dependences in the logarithmic corrections.
Thus, finally we obtain the following scaling relations for the heat transport
\ba
\label{m17}
\Nu\sim{\Pran^{-1/2}\Ra^{1/2}\over(\log\Ra)^2} &&\text{for } \;\Pran\gtrsim1 \;\text{ (regime IV$'_u$)},\qquad\\
\label{m18}
\Nu\sim{\Pran^{1/2}\Ra^{1/2}\over(\log\Ra)^2}  &&\text{for } \;\Pran\lesssim1 \;\text{ (regime IV$'_\ell$)}.\qquad
\ea
The derived scaling relation (\ref{m18})  is the same as in Grossmann and Lohse's model for the ultimate regime \cite{Grossmann2011};
therefore as in that paper we call it ``regime IV$_\ell^\prime$''. 
Relation  (\ref{m17})  gives an extension of that model towards   large $\pr \gtrsim 1$ and we call that regime ``regime IV$_u^\prime$''. 
The derived regimes are sketched in Fig.~\ref{ult}.

The transition between the scaling regimes IV$'_u$ and IV$'_\ell$ takes place at a constant $\Pran$, where $\Nu\sim \Ra^{1/2}/(\log\Ra)^2$ grows slightly slower than $\sim\Ra^{1/2}$ as $\Ra\rightarrow\infty$ (see the  horizontal line for 
$\eta=0$
in the $\Ra-\Pran$ plane in Fig.~\ref{ult}, which indicates the transition 
$\Pran\sim\Ra^\eta$ 
between the neighbouring regimes IV$'_u$ and IV$'_\ell$).
Another boundary for the regime IV$'_\ell$ is for
$\eta=-1$
(marked with a blue line in Fig.~\ref{ult}). 
While moving along this line for an increasing $\Ra$ and $\Pran\sim\Ra^{-1}$, the Nusselt number remains constant, and any steeper transition slope from regime IV$'_\ell$ would imply an unphysical limit $\Nu\rightarrow0$ along that line. 
The blue line in Fig.~\ref{ult} indicates the slope of the transition to the regime II$'_\ell$, which has the very same scaling exponents for $\Nu$ and $\Rey$ as the classical regime II$_\ell$ of the GL-theory \cite{Grossmann2000}. 

Analogously, one can conclude that the slope of the upper boundary of the regime IV$'_u$ should not be steeper than $\Pran\sim\Ra$ so that along this line the Nusselt number remains constant for  increasing $\Ra$.
However, as we will explain below, the transition from regime IV$'_u$ has a significantly more gentle slope, namely 
$\Pran\sim\Ra^{\eta}$ with $\eta=1/3$ 
(marked with a pink line in the sketch of Fig.~\ref{ult}).

Indeed, for the no-slip boundary conditions, each component $u$ of the velocity field vanishes at the (Lipschitz) boundary of the domain, and therefore the Friedrichs inequality $ \lambda_1\langle u^2\rangle \leq \langle (\nn u)^2\rangle $ holds, where $\lambda_1$ is the lowest (positive) eigenvalue of the Laplace operator in the considered domain with the corresponding boundary conditions, that depends only on the geometrical characteristics and has the dimension of inverse squared length \citep{Shishkina2021}.
Therefore, for any RB flow
\be
\label{m20}
\Rey^2\lesssim (L^4/\nu^3) 
\epsilon_u = \Ra\,\Pran^{-2}(\Nu-1),
\ee
where $\Rey^2$ is based on the kinetic energy $\langle u^2 \rangle$. 
In regime IV$'_\ell$ (for small $\Pran$), this relation is always fulfilled within the discussed boundaries, since (\ref{m20}) then means $\Pran\lesssim\Ra$.
However, in regime IV$'_u$, the requirement (\ref{m20}) means $\Pran\lesssim\Ra^{1/3}$, as it follows from the combination of the relations (\ref{m16}), (\ref{m17}) with (\ref{m20}).
Therefore, regime IV$'_u$ can exist only for $\Pran\lesssim\Ra^{1/3}$. 
This is consistent with the 
upper bound of Choffrut {\it et al.} \cite{Choffrut2016}, who, as mentioned above, derived that the upper bounds for the heat transport for large Prandtl numbers $\Pran\gtrsim\Ra^{1/3}$ cannot exceed $\Nu \sim\Ra^{1/3}$ (all up to logarithmic corrections). 

While moving along the line $\Pran\sim\Ra^{1/3}$ with increasing $\Ra$ (the red line in Fig.~\ref{ult}), the Nusselt number effectively scales as $\Nu\sim \Pran^{-1/2}\Ra^{1/2}\sim \Ra^{1/3}$.
We assume that this transition line $\Pran\sim\Ra^{1/3}$ connects regime  IV$'_u$ with regime III$'_\infty$ (as we call it), in which  the scaling exponents are exactly the same as in the classical regime III$_\infty$ of the GL-theory \cite{Grossmann2000}, namely $\Nu\sim \Pran^0\Ra^{1/3}$. 
This result is again consistent with  
the  new
strict upper bound of Choffrut {\it et al.} \cite{Choffrut2016}.

The sketch in Fig.~\ref{ult} summarizes  the  four subregimes of the ultimate regime, namely III$'_\infty$, IV$'_u$, IV$'_\ell$ and II$'_\ell$, which all can be interpreted  as ultimate in the  sense that one can approach infinite $\Ra$ within these regimes.
All these subregimes lie to the right of the pink dotted line that indicates a constant $\Rey_s$ associated with the onset of a turbulent boundary layer.
In the regimes III$'_\infty$ and II$'_\ell$ the scaling exponent $\gamma$ in the relation $\Nu\sim\Ra^{\gamma}$ is, however, smaller than 1/2. 
Regimes IV$'_u$ and IV$'_\ell$ can be considered as the ``true'' ultimate regimes in the sense that only there $\gamma=1/2$.

The proposed model thus indeed suggests that the scaling exponent $\gamma=1/2$ in the scaling relation $\Nu\sim\Ra^{\gamma}$ can be asymptotically achieved 
within the regimes IV$'_u$ and IV$'_\ell$, but only for almost constant Prandtl numbers.
As soon as $\Pran$ changes as a power law  of $\Ra^\xi$ (with some small $|\xi|$, and here it does not matter whether $\xi$ is positive or negative), one should expect an asymptotic reduction of the effective scaling exponent as $\gamma=1/2-|\xi|/2$.

\begin{figure}
\includegraphics[width=0.5\columnwidth]{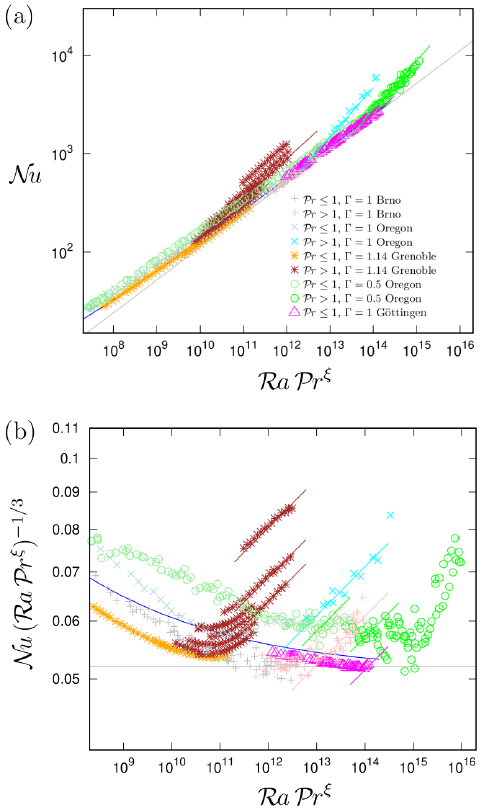}
\caption{
(a) Nusselt number $\Nu$ vs.\ $\Ra\,\Pran^\xi$ (with $\xi=1$ for $\Pran\leq1$ and $\xi=-1$ for $\Pran>1$) and (b) compensated Nusselt number $\Nu\,(\Ra\,\Pran^{\hat \xi})^{-1/3}$ vs.\ $\Ra\,\Pran^{\hat \xi}$ (where the function $\hat \xi(\Pran)\equiv-\tanh(0.5\log_{10}\Pran)$ smoothly connects the two regimes  $\xi=1$ for $\Pran\ll1$ and $\xi=-1$ for $\Pran\gg1$), as obtained in the various different RB experiments of refs.\  \cite{He2012, He2012b, Niemela2000a, Niemela2000, Niemela2000b,Roche2010, Roche2020, Niemela2003, Niemela2003b,Urban2014, Urban2019} under (nearly) Oberbeck--Boussinesq conditions in cylindrical containers, distinguished by the aspect ratio $\Gamma$ and where it was done. 
The blue curve shows the predictions of the GL-theory for the classical regime, $\Pran=1$ and $\Gamma=1$. 
All data sets at the highest achieved Rayleigh numbers show the transition to the ultimate regime, 
\oo
with slopes about $\Nu\sim\Ra^{0.4}$ (brown, cyan, green, pink and magenta thin lines in (b)).
\bb
}
\label{mod2} 
\end{figure}

We now want to compare the available experimental data for large $\Ra$ (close to or in the ultimate regime) with the model Eqs.~(\ref{m17})-(\ref{m18}), cf.~Fig.~\ref{ult}.
According to the model, in the ultimate regime
(for not extremely small or extremely large $\Pran$),
the following scaling should hold: $\Nu\sim\Pran^{\pm1/2}\Ra^{1/2}$, where a negative exponent ($-1/2$) for $\Pran$ should be taken for large $\Pran$ and a  positive one (+1/2) for small $\Pran$.
Thus, the Nusselt number is a function of $\Pran^{\xi}\Ra$ with $\xi=1$ for $\Pran\leq1$ and $\xi=-1$ for $\Pran>1$.
In Fig.~\ref{mod2}a, the considered experimental data for $\Nu$ are plotted vs. $\Pran^{\xi}\Ra$. 
One clearly sees that all data sets, including the Oregon data, follow a scaling close to $\Nu\sim(\Pran^{\xi}\Ra)^{\gamma}$ with $\gamma\approx1/3$ for smaller values of $\Pran^{\xi}\Ra$ and $\gamma$ between 0.4 and 0.5 at the highest values of $\Pran^{\xi}\Ra$.
The exact onset value of the steeper scaling varies from experiment to experiment, which is consistent with the view that 
the transition to the ultimate regime is of  non-normal--nonlinear nature \cite{Roche2020,Lohse2023}. 

For better visibility of the onset, Fig.~\ref{mod2}b provides a compensated plot of $\Nu\,(\Ra\,\Pran^{\hat \xi})^{-1/3}$ vs.\ $\Ra\,\Pran^{\hat \xi}$.
Here, $\hat \xi$ is a function of $\Pran$ that substitutes the discontinuous change of $\xi$ from $+1$ to $-1$ at $\Pran=1$ by a 
smooth function $\hat \xi(\Pran)\equiv-\tanh(d\log_{10}\Pran)$ that matches the small-$\Pran$ and large-$\Pran$ regimes. 
(Here, of course, different options are possible to match the small-$\Pran$ and large-$\Pran$ regimes, in particular, by optimizing the constant $d$, which in Fig.~\ref{mod2}b equals $d=0.5$.)
Again, all data show a transition for very large $\Ra$. 
\oo
The inclined lines in Fig.~\ref{mod2}b highlight the scaling exponent $\gamma=0.4$.
\bb
We interpret the results in Fig.~\ref{mod2}b as support for our new model for the ultimate regime and its subregimes. 

In summary, we have suggested a new model for the heat transfer in  the ultimate regime of RB turbulence, which distinguishes between four subregimes of the ultimate regime and which for each of these subregimes gives the scaling relations as shown in Fig.~\ref{ult}. 
In contrast to prior models, it obeys the 
mathematically strict upper bounds of Choffrut {\it et al.} \cite{Choffrut2016}. 
It moreover is consistent 
with
the experimental data on $\Nu(\Ra, \pr) $ of the various large-$\Ra$ RB experiments of refs.~\cite{He2012, He2012b, Niemela2000a, Niemela2000,Niemela2000b,Roche2010, Roche2020, Niemela2003, Niemela2003b,Urban2014, Urban2019}. 
We emphasize again that in this new representation, which take the $\pr$-dependence into account, the onset of the ultimate regime is seen in all data sets, though at different $\Ra$ numbers, as to be 
expected
for a non-normal--nonlinear instability. 
Our model thus offers a reliable basis to estimate the heat transfer for systems with even larger $\Ra$,  which cannot be achieved in today's experiments, and for 
geophysical and astrophysical systems.

{\it Acknowledgements:} We thankfully acknowledge all colleagues who have made their experimental data available to us and to the whole community. 
 

\end{document}